LAST WORDS ON THE CUPRATES
P W Anderson, Princeton University

Abstract
Much of what I have to say in the following (not all!) is contained in my "informal history" published in 1011 in the Int'l Journ of Mod Phys and available in ArXiv form as Cond.mat/1011.2732. But because the literature on the subject continues to burgeon, mostly without much if any reference to the fundamentals that I laid out in that article, I think it is time to distribute a summary of the conclusions I reached in that article, as well as a few newer thoughts.

In the first place , I have to point out that the "problem", as it is referred to in the naïve discussion which prefaces many of the papers in the literature, is not really a problem: the basic facts of the superconductivity itself, of much of the anomalous behavior which accompanies it, and of the energy (temperature) scale of the phenomena, are not mysterious. The dominant interaction coupling the superconducting electron pairs is the same "superexchange" spin coupling which causes the undoped cuprates to be antiferromagnetic Mott insulators, as I described in my early Science paper in 1987.[1] There I made, perhaps confusingly, the point that the physical nature of the interaction is very similar to that of the ubiquitous valence bond of Paulingesque chemistry, and that it therefore binds singlet pairs of electrons making mobile "valence bonds" between Cu atoms, which become the superconducting carriers.
Within two years, two papers using this interaction and formalisms based on the ideas in the Science paper had presented reasonably quantitative, rough descriptions of "d-wave" superconductivity in the cuprates, one by Rice and collaborators [2]using the renormalized, projected "t-J"

Hamiltonian of Spalek and Rice, and solving it using Monte Carlo; another by Kotliar and a student [3] using a slave Bose field projection technique, essentially equivalent to the t-J method. It is noteworthy that at the time d-wave had not yet been suspected, much less confirmed, so this represented a remarkable prediction, at least five years in advance of its experimental confirmation. A feature of that prediction worth noting is that it unequivocally led to a **real energy gap vanishing at nodes in k-space**, inexplicable in other terms. This arises because of a mechanism rather unique to this system, which I have called "spin-charge locking".[4]

A fairly complete exposition of the mean field theory was provided by Edegger, Gros and others[5] in 2006.

I would emphasize that the pairing interaction is not appropriately described by the term "spin fluctuations", as it was in work by Scalapino, Pines, and others. "Spin fluctuations" is a term originated by Doniach, Schrieffer et al to describe the (possibly) long-range fermion-fermion interactions near a **ferromagnetic critical** point and can be logically viewed as an extension of a Feynman diagrammatic theory for an almost ferromagnetic Fermi liquid. Antiferromagnetic superexchange[6] is **not** derivable from a simple Feynman theory [7] and has no special relation to fluctuations near a critical point, and the numerous attempts to fit parameters to such a theory have led only to confusion. Also, the interaction is demonstrably short range in space and time.

This confusion of antiferromagnetic "kinetic" exchange with ferromagnetic "direct" exchange has afflicted generations of researchers. In particular, full-scale numerical methods indeed eventually led to d-wave superconductivity,[8] because they were essentially including all of the relevant physics; but when reinterpreted in terms of Feynman graph concepts such as vertices and self-energies they gave misleading answers. Analytic methods are essential in this case if one is looking for an effective low-energy theory with which to predict the range of material properties, not just the ground state. When the experimental observations demonstrate that the behavior is not conventional Eliashberg, it is not useful to fit Eliashberg parameters to numerical calculations.

Conventional Fermi liquid theory is necessarily symmetrical between electrons and holes in the neighborhood of the Fermi surface, and the striking and ubiquitous asymmetry of the Green's functions of electrons vs holes which is revealed by vacuum tunneling experiments (and which follows from projective theories) is the most prominent experimental effect which demonstrates the inappropriateness of conventional theory. I remain baffled by the almost universal refusal of theorists to confront this evident fact of hole-particle asymmetry head on. Its meaning is that the first step of any attempt at theory must be the perturbative "Gutzwiller"-type projective transformation, first formalized by Kohn,[9] to a Hamiltonian with matrix elements to double occupancy of holes eliminated. This has the important effect of allowing a consistent treatment of kinetic exchange as the consequence of the projection.

I should give prominence to a series of papers by Gros, Edegger, Muthukumar and others[10] who worked out further numerical consequences of the skeletal theory sketched above and showed that it does indeed fit the experimental facts in some detail.

A second fact which was discovered in the early literature and should be kept in mind throughout is the severe limitation of the superconducting Tc by thermal excitation of vortex fluctuations. It was early remarked that superconducting fluctuations were expected to be the strongest limitation on Tc, by PWA (see ref 1) and Baskaran in 1987 and by Fukuyama shortly thereafter; but the community as a whole rightly paid more attention to Emery and Kivelson,[11] who showed quantitatively in 1995 that the universal linear relationship between Tc and $\rho_s$, the superfluid density, which had been emphasized by Uemura, could be interpreted as meaning that Tc is a "BKT" vortex proliferation transition in the thermally decoupled planar square CuO2 lattices. That this relationship holds up to and beyond the maximum Tc of the "dome" shows unequivocally that this is the fundamental limitation on Tc for the superexchange mechanism. The renormalization of $\rho_s$ by projection is, both experimentally and theoretically, proportional to the ratio g=2x/(1+x), where x is the doping fraction, and therefore Tc drops drastically at small x. The effective antiferromagnetic exchange integral renormalizes downward with increasing x, as I have estimated[12] using methods similar to those of Brinkman and Rice, leading to the decrease in Tc at high doping and the familiar dome-shaped x-dependence of Tc.

Although it is not a fact generally remarked in the textbooks, the BKT transition is not from a superconductor into a normal metal. The phase above the transition as formally described in the BKT theory actually has no mechanism for destroying the superconducting pair field, only for disordering its phase, so that formally a space-time-dependent $\rho_s(r,t)$ is present above Tc, with, near Tc, the same magnitude as below Tc but without long-range phase stiffness. The kinetic energy of flow of this fluid may be written, for the two-dimensional case, as the sum of two terms. The first is the sum, over all pairs of vortices, of the logarithmic interactions

$$E_{ij}=q_iq_j(h^2\rho_s/2m)\ln(r_{ij}/a)$$

between each pair of vortices. Here q is the sign of the circulation of vortex i, and a is a core radius; $r_{ij}$ is the distance between each pair ij of vortices. This is the term which is used in the conventional BKT theory and which allows the proliferation of vortex pairs above Tc, which gives the metal resistance. But the density of vortex pairs does not immediately diverge above Tc , it rises(according to theory) as $\exp-(2/\sqrt{(T-T_c)})$, thus does not even have a singular derivative at Tc. Therefore for a considerable range of T above Tc we may assume that the superfluid continues to be well represented as a liquid of vortices, and this is the "Ong vortex liquid". It terminates (in T) at a higher transition whose nature I do not know.

The second term which arises from the kinetic energy of the vortex array is, interestingly, independent of the positions $r_i$ of the individual vortices, depending only on the total vorticity $\Sigma_i q_i$ and diverging with the upper cutoff size R of the sample:
$$E_2=(\Sigma_i q_i)^2(h^2\rho_s/m)\ln(R/a)$$
In the standard discussions of the BKT transition this divergent term is removed by assuming that the sum of the q's is 0; but it is not remarked that this is equivalent to the statement that in zero field, even above Tc the system retains a *topological* order: the net circulation measured along any large loop vanishes. But this is true, of course, only in the absence of a magnetic field; the correct prescription for removing this divergence in a field must be that the sum of the q's must match the net density of flux quanta in the field, as was pointed out decades ago by Onsager, Feynman and Abrikosov.
This term is manifestly unaffected by the specific locations of the vortices: **it manifestly cannot be screened out by any rearrangement of the thermally excited vortices in the liquid!** This somewhat counterintuitive result was not emphasized in the textbook treatments of BKT systems and is responsible for the striking diamagnetism and Nernst effect of these vortex liquids,. which was first observed in 2000 and explored in detail by Wang, Li and Ong . [13]

The effect of a uniform density of unpaired vortices of the same sign is to mimic the large-scale flow caused by a uniform field at any scale larger than the distance between the unpaired vortices, which may be taken to be roughly $R_B = \sqrt{(hc/2eB)}$.  Our argument is that this flow energy can in fact be screened out by a uniform density of unpaired vortices at scales larger than $R_B$, but that the granularity of the flow caused by the quantization of vorticity cannot be neglected at scales less that $R_B$, so that every field vortex (with density proportional to B) contributes a self-energy proportional to $\ln(R_B/a)$ or $\frac{1}{2} \ln (1/B)$.  This BlnB term (B for the number of vortices) is the characteristic BlnB dependence which appears in the Nernst and diamagnetism measurements on all underdoped cuprates, and which is so strikingly continuous at finite field in the field-temperature plane as one passes the nominal Tc.  This is so even though the superconductivity is 3-dimensional below Tc, but 2-dimensional above.

A great deal of attention has been devoted to the quantum oscillations that appear in some of the cuprates at intermediate doping levels and high fields.  There is in particular a transition into a state which appears to result from a CDW formation which brings into existence a 3-dimensional Fermi surface pocket, whose nature has been investigated most thoroughly in the "ortho-2" doping of YBCO. It now appears, from extensive investigations by Ong among others,[14] that these phenomena occur only above a large threshold field of around 18 Tesla which destroys the smaller energy gaps around the gap nodes, permitting the Fermi surface pockets to grow into Fermi arcs and to reconnect (as described by Sebastian and Harrison) by means of scattering on a lattice distortion—basically, a CDW-induced magnetic breakdown. (or, equivalently, one can think of it as magnetic breadown inducing a CDW.) The whole phenomenon leaves intact the large antinodal gaps which disappear only at much higher fields, and which determine $H_{c2}$. The phase diagram showing this is reproduced here in a figure from reference 14.

The region of the phase diagram which is best understood theoretically is the region to the right of and above the pairing transition, which is commonly called T*. This is where the characteristic symptoms of the "strange metal" occur: a Hall constant rising at low T, a "linear-T" resistivity and, more crucially, a power law Drude tail in the infrared conductivity, as established by van der Marel[15] (following Schlesinger and Bontemps.) This is the region which may be thought of as a continuation from the "t-model", where the superexchange interaction J, of order $t^2/U$, may be treated as a small perturbation relative to the renormalized kinetic energy tx. All of the phenomena are caused by the Kohn-Spalek-Rice projective transformation eliminating matrix elements to double occupancy of holes.

The formal discussion of the large U case has never been written down explicitly but it goes more or less like this:
As we increase U towards the critical value Uc at which the ladder diagrams diverge, at every U there is a Luttinger-size Fermi surface of zero-energy Fermionic excitations which decay at a rate proportional to $\omega^2=(E-E_F)^2$, and are therefore exact at a Fermi surface.
It is at least possible that this Fermi surface of exact excitations, with Luttinger volume, continues to exist through and above Uc.

This is, in fact, the basic Ansatz of the Hidden Fermi Liquid theory: that the Hilbert space of the low-energy excitations in the states *above* Uc is representable in terms of these excitations, but that this Hilbert space must be restricted by projecting out double occupancy, because of the divergence of the ladder diagrams which leads to the "upper Hubbard band" of doubly-occupied states. These therefore must not be counted as part of the Hilbert space allowed to the excitations near the Fermi level.

We represent the Fermionic excitations by a band of renormalized pseudoparticles $c_k^*$, $c_k$, and the Fourier transformed local operators by $c_i^*$, $c_i$. Then the Gutzwiller-like projection process which eliminates double occupancy may be carried out by multiplying, in the band kinetic energy, every $c_{i,\sigma}$ and $c^*_{i,\sigma}$ by the projector $P = c_{i,-\sigma} c^*_{i,-\sigma}$. The mean field theory used by Rice et al, and effectively by Kotliar et al, and later by Paramekanti et al, and Edegger et al, involves replacing the projectors by their mean values (and renormalizing, incorrectly in the early papers as shown by Pretko[16], the size of the exchange interaction from that given by perturbation theory). The Hidden Fermi liquid theory instead observes that in the normal Fermi liquid the Green's function of the projector has an algebraic singularity which vanishes at $\omega=0$, $q=0$, and the projector does not have a true mean value, because of the entanglement properties of the Fermi sea ground state. This is the same type of algebraic singularity that enters in the "x-ray edge" singularity (the 'orthogonality catastrophe") of normal metals. It is related to the "chiral anomaly" of field theory, in that it is caused by the deep entanglement of the Fermion fluid, and like it appears as a term which affects the spectrum at all energies. The HFL theory takes the simplest path of calculating the Green's function of the composite operator as the product of those of its two components.

As we move into the underdoped regime an additional complication arises. The down-spin projector multiplying the up-spin particle has a component which can be thought of as a down-spin particle $c_{i,-\sigma}^*$ multiplying a spin operator $S_i^+$. This will cause structure in the tunneling spectrum, especially at the frequency of the "soft mode" resonance at several tens of mev which can be (and is) misinterpreted as indicating a role in causation of the superconducting gap.

The anomaly becomes much less severe in the optimally doped .superconductor because the energy gap allows the infrared end to converge, and the mean field theory should be otherwise quite satisfactory in this case, as shown in the fits by Anderson and Ong [17] of a projected BCS theory to tunneling data. The tunnelling data also seem to confirm that there is an actual zero of the energy gap, again confirming that it is real rather than complex.

There remain a number of details in the phase diagram which have not yet become clear to me. I have mentioned the upper temperature limit of the Ong vortex liquid phase, which seems rather sharp and leads to a phase which still shows some evidence of pairing. I also do not pretend to understand the transport theory in this region; I have given arguments that $h/\tau \sim T$ but they are very qualitative.

But most persistently, there appears evidence of some kind of T-symmetry breaking, and recently of second harmonic generation, near the T* line where one presumes pairing first appears. One should emphasize that the incontrovertible evidence for gap zeroes at low temperatures from thermal conductivity and thermal Hall conductivity measurements[18] proves that the well-developed energy gap is real, the gap zeros are protected, and hence demonstrates T-invariance in the ground state. But since the process of spin-charge locking is due to energy terms which are not relevant at the pairing transition, it may be that the pairing transition is more complicated than we envision and leaves room for a complex gap phase ("d+id") near T*. Otherwise I must confess to my failure to find a reasonable explanation for these observations.

Let me, in closing, say a few words about problem-solving strategy in materials science. In the first place, one must realize that unlike particle physics or cosmology the problems in materials science are over- rather than under-determined. There is much too much data. Any hypothesis which does not contradict something evident in the data is likely to be true. Equally, it will usually be easy to exclude false or irrelevant ideas.

For instance, one set of evident facts is that in reasonably well-developed superconducting states, the effective mass is low, there are well-developed and persistent zeroes of the gap(hence it is real),and the transition out of superconductivity is caused by phase disorder. Thus bipolarons play no central role, nor do loop currents and T asymmetry(which is puzzling but peripheral); the superfluid density is renormalized downwards linearly in x but not otherwise changed. Second, the simplest theory produces an obviously satisfactory explanation of the scale and central nature of the phenomena—d-wave, $t^2$/U, rapid decrease of scale as doping is increased, asymmetric Green's functions. It is not surprising that at low doping, where the kinetic energy has been drastically reduced, some compositions will exhibit inhomogeneous phases, but there is no evidence that the central phenomenon of superconductivity with a gap in the region of tens of mev is profoundly dependent on such details; they mostly tend to compete with it. All the fuss about a

"mystery" seems almost a fiction providing gainful employment for theorists.

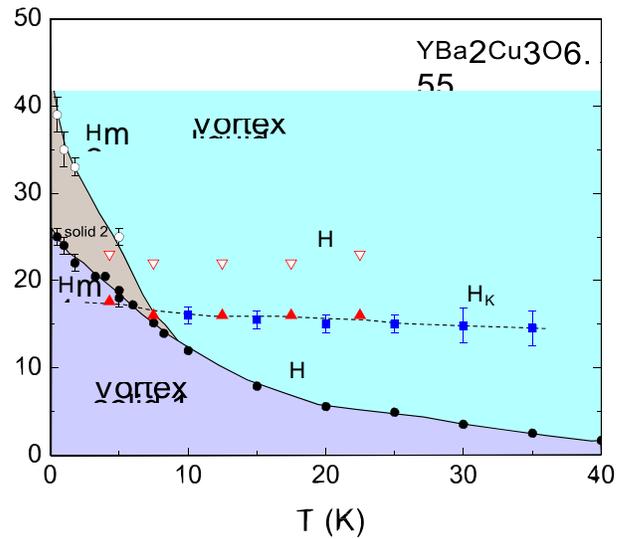

Fig. 8. Magnetic phase diagram in untwinned YBa$_2$Cu$_3$O$_y$ ($y$ = 6.55) inferred from magnetization and the thermal conductivity. Given the 2D nature of the transition curves, the vertical axis refers to the $z$ component of H applied in the torque experiments (for $\kappa_{xx}$ measurements, H is always k$\hat{z}$). Above ~ 8 K, the vortex solid is stable below the melting field $H_m$ (solid circles). Below ~ 8 K, the $H_m$ curve splits into two branches: $H_{m1}$ (solid circles) and $H_{m2}$ (open circles). The vortex solid 1 below $H_{m1}$ displays a large critical current density $J_c$ and shear modulus, whereas the solid 2 ($H_{m1} < H < H_{m2}$) has a much smaller $J_c$ but survives to ~ 41 T at 0.5 K. Throughout, the vortex- liquid state is stable to at least 41 T (region shaded light blue), but likely much higher judging from the trend in $M_d$ vs. $H$. The nearly $T$-independent kink field $H_K$, identified as the onset of static charge order, is plotted as blue squares (inferred from $\chi_d$) and as red solid triangles (from $\kappa_a$ and $\kappa_b$). It in- tersects $H_{m1}$ without affecting the melting (to our resolution). The field scale $H_p$ (open triangles, derived from $\kappa_a$ and $\kappa_b$) appears to terminate at the lower melting field $H_{m1}$ in the limit $T \rightarrow 0$. At 0.5 K, the dHvA oscillations onset near 25 T (coexist with the vortex liquid below ~ 41 T).